\documentclass[english]{article}
\usepackage{lmodern}
\usepackage[T1]{fontenc}
\usepackage[latin9]{inputenc}
\usepackage{geometry}
\usepackage{graphicx}
\usepackage{setspace}
\usepackage{esint}
\usepackage[authoryear]{natbib}
\usepackage{subscript}
\makeatletter
\newcommand{\lyxaddress}[1]{
\par {\raggedright #1
\vspace{1.4em}
\noindent\par}
}

\@ifundefined{date}{}{\date{}}
\makeatother

\usepackage{babel}
\begin{document}

\title{Factors controlling the time-delay between peak CO\textsubscript{2}
emissions and concentrations}

\author{Ashwin K Seshadri}

\maketitle

\lyxaddress{Divecha Centre for Climate Change, Indian Institute of Science, Bangalore
560012. India. (ashwin@fastmail.fm)}

\pagebreak{}
\begin{abstract}
Carbon-dioxide (CO\textsubscript{2}) is the main contributor to anthropogenic
global warming, and the timing of its peak concentration in the atmosphere
is likely to govern the timing of maximum radiative forcing. It is
well-known that dynamics of atmospheric CO\textsubscript{2} is governed
by multiple time-constants, and here we approximate the solutions
to a linear model of atmospheric CO\textsubscript{2} dynamics with
four time-constants to identify factors governing the time-delay between
peaks in CO\textsubscript{2} emissions and concentrations, and therefore
the timing of the concentration peak. The main factor affecting this
time-delay is the ratio of the rate of change of emissions during
its increasing and decreasing phases. If this ratio is large in magnitude
then the time-delay between peak emissions and concentrations is large.
Therefore it is important to limit the magnitude of this ratio through
mitigation, in order to achieve an early peak in CO\textsubscript{2}
concentrations. This can be achieved with an early global emissions
peak, combined with rapid decarbonization of economic activity, because
the delay between peak emissions and concentrations is affected by
the time-scale with which decarbonization occurs. Of course, for limiting
the magnitude of peak concentrations it is also important to limit
the magnitude of emissions throughout its trajectory, but that aspect
has been studied elsewhere and is not examined here. The carbon cycle
parameters affecting the timing of the concentration peak are primarily
the long multi-century time-constant of atmospheric CO\textsubscript{2},
and the ratio of contributions to the impulse response function of
atmospheric CO\textsubscript{2} from the infinite time-constant and
the long time-constant respectively. Reducing uncertainties in these
parameters can reduce uncertainty in forecasts of the radiative forcing
peak.
\end{abstract}

\section*{Keywords}

Global climate change; carbon dioxide; peak radiative forcing; climate
change mitigation; decarbonization.

\section{Introduction}

As countries agree on commitments towards a new international climate
treaty to be decided in 2015 (\citet{UNFCCC2014,UNFCCC2014a}), these
will include mitigation of not only carbon-dioxide (CO\textsubscript{2})
but also other climate forcers (\citet{UNFCCC2014b}). CO\textsubscript{2}
is, and is likely to remain, the largest contribution to radiative
forcing (\citet{Forster2007,Myhre2013}). Limiting long-term warming
requires limiting the growth in global CO\textsubscript{2} emissions,
and eventually reducing these emissions. If the present increasing
trend in global CO\textsubscript{2 }emissions is eventually reversed
so that an emissions peak occurs, the corresponding peak in concentration
will be delayed because of its long atmospheric lifetime (\citet{Allen2009,Meinshausen2009,Mignone2008}).
A CO\textsubscript{2} concentration peak would be a significant event
for global climate: it would govern the maximum contribution of CO\textsubscript{2}
emissions to radiative forcing. Furthermore, assuming that CO\textsubscript{2}
continues to be the major contribution to radiative forcing, then
its peak concentration will strongly influence the magnitude and timing
of peak global warming.

The Earth's CO\textsubscript{2} cycle is complex, involving multiple
reservoirs that maintain exchanges occurring at very different rates
(\citet{Archer1997,Cox2000,Falkowski2000}). The most rapid uptake
of excess CO\textsubscript{2} is by the surface ocean and land biosphere
(\citet{Pierrehumbert2014}). Progressively slower processes involve
mixing with the deep-ocean, reduction of ocean acidity due to dissolution
of carbonates, and uptake of excess atmospheric CO\textsubscript{2}
via reaction with CaCO\textsubscript{3} or silicate rocks on land
(\citet{Archer1997,Archer2008,Archer2009}). The last two processes
require many tens of thousands of years so that, on the timescales
of the next few centuries, their contributions can be effectively
neglected. Equivalently their effects can be treated as occurring
with infinite time-constant. Accurate characterization of the different
processes involved, in order to describe the fate of excess atmospheric
CO\textsubscript{2}, requires coupled climate-carbon-cycle or Earth-system
models; such models have been employed to describe effects of mitigation
scenarios on CO\textsubscript{2} in the atmosphere (\citet{Petoukhov2005,Friedlingstein2006}).
As the mitigation of CO\textsubscript{2} emissions unfolds, these
and similar models will play important roles in estimating the consequences
for atmospheric CO\textsubscript{2}, including the timing and magnitude
of its peak concentration.

This paper solves a linear model of atmospheric CO\textsubscript{2}
with four time-constants (\citet{Joos2013}) to understand the factors
controlling the time-delay between peaks in emission and concentration
and therefore the timing of the concentration peak. Previous studies
have described the relationship between mitigation and warming, and
highlighted the importance of rapid mitigation (for e.g. \citet{Socolow2007,Allen2009,Allen2014,Huntingford2012}).
Here we focus specifically on solving the model of atmospheric CO\textsubscript{2}
analytically, to identify some of the important factors controlling
the time to the concentration peak of CO\textsubscript{2}.

\section{Models of emissions and carbon cycle}

\subsection{Carbon cycle model}

\citet{Joos2013} estimated the impulse response of CO\textsubscript{2}
in the Earth's atmosphere, by averaging a group of Earth System Models.
They estimated a mean response with four time-constants
\begin{equation}
h\left(t\right)=0.276e^{-t/4.30}+0.282e^{-t/36.5}+0.224e^{-t/394}+0.217\label{eq:pn1}
\end{equation}
which we apply, and compute atmospheric concentration using
\begin{equation}
u\left(t\right)=\int_{0}^{t}m\left(z\right)h\left(t-z\right)dz+u_{P}\label{eq:pn2}
\end{equation}
where $m\left(z\right)$ is anthropogenic emissions in concentration
units, i.e. the rate of increase in concentration in the hypothetical
case of infinite atmospheric lifetime. The constant term $u_{P}$
is preindustrial concentration. Concentration is noted in parts per
million (ppm). Atmospheric emission of CO\textsubscript{2} in the
year 2013 was $36\times10^{12}$ kg, equivalent to $4.5$ ppm.%
\footnote{If CO\textsubscript{2} had infinite lifetime, emissions in the year
2013 would have increased atmospheric concentration by $4.5$ ppm. %
} We furthermore describe the impulse response function symbolically
as
\begin{equation}
h\left(t\right)=\sum_{i=1}^{K}\mu_{i}e^{-t/\tau_{i}}\label{eq:pn3}
\end{equation}
 with $\sum_{i=1}^{K}\mu_{i}=1$ and $\left\{ \tau_{1},\tau_{2},\tau_{3},\tau_{4}\right\} =\left\{ 4.4,36.5,394,\infty\right\} $
following the results of \citet{Joos2013}. The infinite time-constant
approximates the effects of very slow processes involving buffering
of ocean acidity by dissolution of carbonates and the uptake of CO\textsubscript{2}
in the weathering of rocks. 

There is uncertainty in the function $h\left(t\right)$, with different
earth system models likely to yield different results (\citet{Joos2013}).
While the present paper does not characterize this uncertainty, Section
3.3 considers the influence of changing those parameters in this 4-time-constant
model that are shown to affect the timing of the concentration peak.

\subsection{Emissions model}

The model of emissions $m\left(t\right)$ is very simple, and chosen
so as to describe emissions using a few different parameters that
can be readily interpreted. The emissions model is $m\left(t\right)=m_{0}\left(1+r\right)^{\left(min\left(t,t_{g}\right)-t_{0}\right)}e^{-\left(t-t_{0}\right)/\tau_{m}}$
(\citet{Seshadri2015a,Seshadri2015b}), with $m_{0}$ being present
emissions, $r$ the growth rate of gross global product (GGP), and
$t_{0}$ denoting the present time. It is assumed, for the purposes
of the emissions model used here, that GGP increases for $t_{g}$
years from the present at constant rate $r$, after which it remains
fixed. The term $e^{-\left(t-t_{0}\right)/\tau_{m}}$ describes the
effect of decrease in emissions intensity of economic output, with
$\tau_{m}\rightarrow\infty$ corresponding to the absence of any mitigation,
and smaller values of $\tau_{m}$ corresponding to rapid mitigation.
This model of emissions has been borrowed from previous work (\citet{Seshadri2015a,Seshadri2015b}).

\section{Results}

\subsection{Peak atmospheric concentration of CO2}

With emissions $m\left(t\right)$ the atmospheric concentration is
$u\left(t\right)=u_{P}+\int_{0}^{t}m\left(z\right)h\left(t-z\right)dz$,
and differentiating this equation we obtain for the rate of change
of concentration that $u'\left(t\right)=m\left(t\right)+\int_{0}^{t}m\left(z\right)\frac{\partial h\left(t-z\right)}{\partial t}dz$,
where we have used the fact that $h\left(0\right)=1$ from equation
(\ref{eq:pn1}). Using equation (\ref{eq:pn3}) we obtain that $u'\left(t\right)=m\left(t\right)-\sum_{i=1}^{K}\frac{\mu_{i}}{\tau_{i}}e^{-t/\tau_{i}}\int_{0}^{t}e^{z/\tau_{i}}m\left(z\right)dz$.
The last integral can be evaluated by parts to give $\int_{0}^{t}e^{z/\tau_{i}}m\left(z\right)dz=\tau_{i}m\left(t\right)e^{t/\tau_{i}}-\tau_{i}\int_{0}^{t}e^{z/\tau_{i}}m'\left(z\right)dz$,
and substituting this yields 
\begin{equation}
u'\left(t\right)=\sum_{i=1}^{K}\mu_{i}e^{-t/\tau_{i}}\int_{0}^{t}e^{z/\tau_{i}}m'\left(z\right)dz\label{eq:pn4}
\end{equation}
for the rate of change of concentration. This result has been derived
previously in \citet{Seshadri2015}. 

For short time-constants $\tau_{1}=4.4$ years and $\tau_{2}=36.5$
years for which $t\gg\tau_{i}$, $i=1,2$, we can approximate the
integral $\int_{0}^{t}e^{z/\tau_{i}}m'\left(z\right)dz$ by $\tau_{i}e^{t/\tau_{i}}m'\left(t\right)$
. For the very long time-constant that is represented by $\tau_{4}=\infty$
the integral becomes $\int_{0}^{t}e^{z/\tau_{i}}m'\left(z\right)dz=\int_{0}^{t}m'\left(z\right)dz=m\left(t\right)$
. Hence the rate of change of concentration becomes approximately
\begin{equation}
u'\left(t\right)\cong\left(\mu_{1}\tau_{1}+\mu_{2}\tau_{2}\right)m'\left(t\right)+\mu_{3}e^{-t/\tau_{3}}\int_{0}^{t}e^{z/\tau_{3}}m'\left(z\right)dz+\mu_{4}m\left(t\right)\label{eq:pn5}
\end{equation}

We denote the time to the emissions peak as $t_{1}$, the time to
the concentration peak as $t_{2}$, and the time-delay as $\delta t\equiv t_{2}-t_{1}$.
Then the integral in equation (\ref{eq:pn5}) above can be described
as the sum of integrals from $0$ to $t_{1}$, and from $t_{1}$ to
$t_{2}$. Furthermore, defining weighted rates of change of emissions
\begin{equation}
m'_{av,i}=\frac{\int_{0}^{t_{1}}e^{z/\tau_{3}}m'\left(z\right)dz}{\int_{0}^{t_{1}}e^{z/\tau_{3}}dz}\label{eq:pn6}
\end{equation}
and 
\begin{equation}
m'_{av,d}=\frac{\int_{t_{1}}^{t_{2}}e^{z/\tau_{3}}m'\left(z\right)dz}{\int_{t_{1}}^{t_{2}}e^{z/\tau_{3}}dz}\label{eq:pn7}
\end{equation}
we obtain the formula for the rate of change of concentration at time
$t=t_{2}$ 
\begin{equation}
u'\left(t_{2}\right)\cong\left(\mu_{1}\tau_{1}+\mu_{2}\tau_{2}\right)m'\left(t_{2}\right)+\mu_{3}\tau_{3}m'_{av,i}e^{-t_{2}/\tau_{3}}\left(e^{t_{1}/\tau_{3}}-1\right)+\mu_{3}\tau_{3}m'_{av,d}e^{-t_{2}/\tau_{3}}\left(e^{t_{2}/\tau_{3}}-e^{t_{1}/\tau_{3}}\right)+\mu_{4}m\left(t_{2}\right)\label{eq:pn8}
\end{equation}

Before proceeding we identify the factors influencing whether the
concentration will reach a peak value and eventually decline, as opposed
to continuing to increase to an asymptotic maximum. Consider emissions
scenarios where the emissions peaks and then declines to zero so that,
eventually, $m'\left(t\right)=0$ and $m\left(t\right)=0$. In that
case only the large but finite time-constant $\tau_{3}=394$ years
in the model of \citet{Joos2013} plays a role in the sign of $u'\left(t\right)$.
The value of this quantity then becomes, for $t>t_{1}$, approximately
\begin{equation}
u'\left(t\right)\cong\mu_{3}\tau_{3}m'_{av,i}e^{-t/\tau_{3}}\left(e^{t_{1}/\tau_{3}}-1\right)+\mu_{3}\tau_{3}m'_{av,d}e^{-t/\tau_{3}}\left(e^{t/\tau_{3}}-e^{t_{1}/\tau_{3}}\right)\label{eq:pn9}
\end{equation}
which is negative if 
\begin{equation}
-m'_{av,d}>m'_{av,i}\frac{e^{t_{1}/\tau_{3}}-1}{e^{t/\tau_{3}}-e^{t_{1}/\tau_{3}}}\label{eq:pn10}
\end{equation}
requiring the rate of decrease of emissions to be sufficiently large
in magnitude. The larger the average rate of increase in emissions,
and the longer the increase persists, the more stringent is the condition
on the subsequent decrease in order for concentrations to eventually
decrease. Therefore an early peak in global emissions can help stabilize
atmospheric CO\textsubscript{2} concentrations, as would be expected. 

We now seek an expression for the time $t_{2}$ to the CO\textsubscript{2}
concentration peak, assuming that mitigation occurs rapidly enough
for one to occur. The condition for such a peak is that the rate of
change of concentration vanishes. Hence, from equation (\ref{eq:pn8})
above 
\begin{equation}
u'\left(t_{2}\right)\cong\left(\mu_{1}\tau_{1}+\mu_{2}\tau_{2}\right)m'\left(t_{2}\right)+\mu_{3}\tau_{3}m'_{av,i}e^{-t_{2}/\tau_{3}}\left(e^{t_{1}/\tau_{3}}-1\right)+\mu_{3}\tau_{3}m'_{av,d}e^{-t_{2}/\tau_{3}}\left(e^{t_{2}/\tau_{3}}-e^{t_{1}/\tau_{3}}\right)+\mu_{4}m\left(t_{2}\right)=0\label{eq:pn11}
\end{equation}
and, writing $m\left(t_{2}\right)=m'_{i}t_{1}+m'_{d}\left(t_{2}-t_{1}\right)$,
where $m'_{i}$ and $m'_{d}$ are average rates of change during emission\textquoteright s
increasing and decreasing phases respectively, and collecting terms
involving $t_{2}$ yields
\begin{equation}
\mu_{3}\tau_{3}e^{-t_{2}/\tau_{3}}\left(\left(m'_{av,i}-m'_{av,d}\right)e^{t_{1}/\tau_{3}}-m'_{av,i}\right)+\mu_{4}m'_{d}t_{2}\cong-\mu_{3}\tau_{3}m'_{av,d}-\mu_{4}\left(m'_{i}-m'_{d}\right)t_{1}-\left(\mu_{1}\tau_{1}+\mu_{2}\tau_{2}\right)m'\left(t_{2}\right)\label{eq:pn12}
\end{equation}
We can neglect the last term on the right side of equation (\ref{eq:pn12}),
since $m'\left(t_{2}\right)$ is comparable in magnitude to $m'_{av,d}$,
as will be shown, but $\mu_{1}\tau_{1}+\mu_{2}\tau_{2}\ll\mu_{3}\tau_{3}$.
Hence the time $t_{2}$ to the concentration peak in the model is
governed by approximate equality 
\begin{equation}
\mu_{3}\tau_{3}e^{-t_{2}/\tau_{3}}\left(m'_{av,i}-\left(m'_{av,i}-m'_{av,d}\right)e^{t_{1}/\tau_{3}}\right)-\mu_{4}m'_{d}t_{2}\cong\mu_{3}\tau_{3}m'_{av,d}+\mu_{4}\left(m'_{i}-m'_{d}\right)t_{1}\label{eq:pn13}
\end{equation}
The approximate equality in equation (\ref{eq:pn13}) has been verified
in Figure 1c, for emissions scenarios shown in Figure 1a and corresponding
concentration graphs plotted in Figure 1b. It is therefore confirmed
that the short time-constants can be approximately neglected while
studying the concentration peak. The above equation has to be solved
numerically because neither the exponential nor the linear term in
$t_{2}$ on the left side of the expression can be neglected. That
this is the case is shown in Figure 1d, which plots these two terms.
Neither term is dominant. 

However in order to understand qualitatively the factors influencing
the time $t_{2}$, let us imagine that $t_{2}$ is so large that the
second term in the left side of equation (\ref{eq:pn13}), $-\mu_{4}m'_{d}t_{2}$,
were dominant. Then the solution would be given by 
\begin{equation}
t_{2}\cong\tau_{3}\left(\left(\frac{m'_{i}}{-m'_{d}}+1\right)\frac{t_{1}}{\tau_{3}}+\frac{\mu_{3}}{\mu_{4}}\frac{m'_{av,d}}{-m'_{d}}\right)\label{eq:pn14}
\end{equation}
The time to the concentration peak increases with the rate of increase
of emissions during their growing phase. It decreases with the rate
of decrease of emissions during their declining phase. The time to
the concentration peak increases with the ratio $\mu_{4}/\mu_{3}$,
describing the ratio of the impulse response of atmospheric CO\textsubscript{2}
coming from the infinite time-constant $\tau_{4}$ and finite but
long time-constant $\tau_{3}$ respectively. It increases with the
time $t_{1}$ to the emissions peak, and in fact the influence of
$t_{1}$ can be described in terms of dimensionless ratio $t_{1}/\tau_{3}$.
However the left side of equation (\ref{eq:pn13}) increases with
$t_{1}$, because of the exponential term, so the influence of $t_{1}$
is not as strong as equation (\ref{eq:pn14}), which neglects the
influence of this term, would suggest. Similarly the influence of
$\tau_{3}$ is not as strong as equation (\ref{eq:pn14}) would suggest
because the exponential term in equation (\ref{eq:pn13}) also increases
with this quantity. 

\begin{figure}
\includegraphics{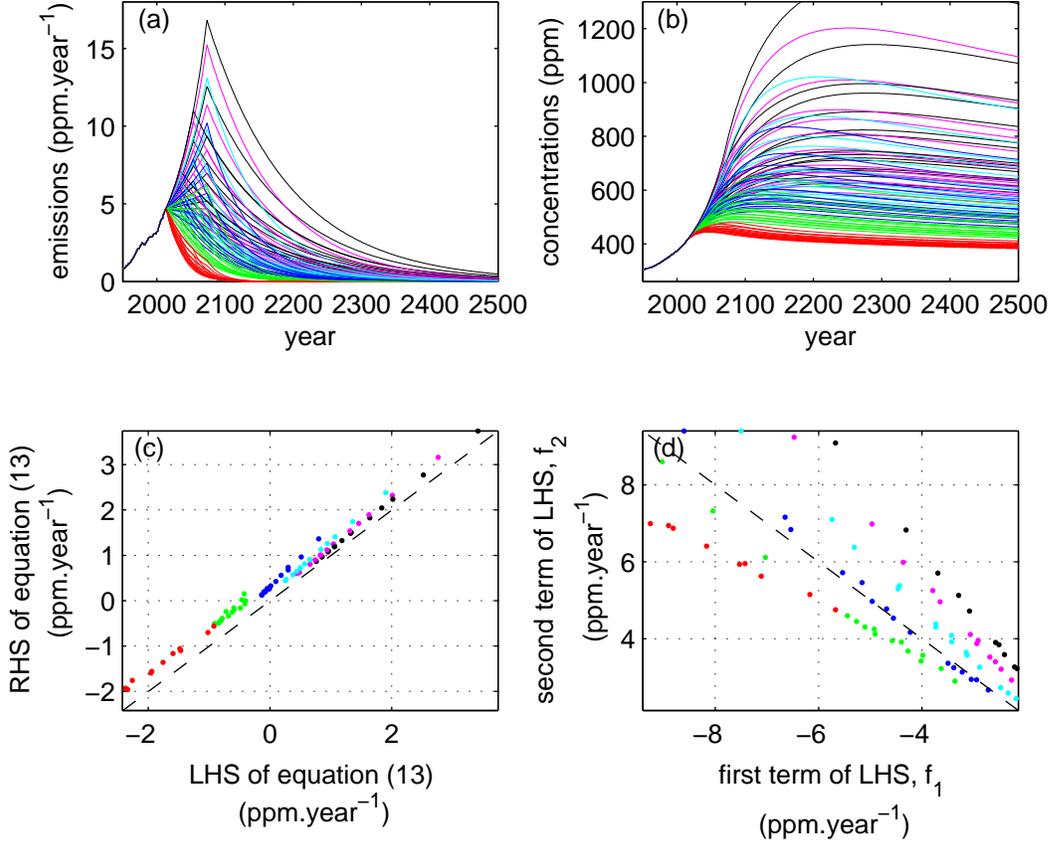}

\protect\caption{Verification of approximation in equation (\ref{eq:pn13}) and demonstration
that neither term on the left side of this equation is dominant for
a wide range of CO\protect\textsubscript{2} emissions scenarios:
(a) emissions pathways; (b) corresponding graphs of CO\protect\textsubscript{2}
concentration; (c) each side of equation (13), with left side in abscissa
and right side in ordinate. The dashed line shows where abscissa and
ordinate are equal; (d) first and second terms on the left side of
equation (\ref{eq:pn13}), $f_{1}=\mu_{3}\tau_{3}e^{-t_{2}/\tau_{3}}\left(m'_{av,i}-\left(m'_{av,i}-m'_{av,d}\right)e^{t_{1}/\tau_{3}}\right)$
and $f_{2}=-\mu_{4}m'_{d}t_{2}$ , showing that neither term is dominant.
Dashed line shows were abscissa and ordinate are of equal magnitude.
Colors in panels correspond to different e-folding mitigation timescales
in the respective emissions scenarios (see legend in Figure 2). }

\end{figure}

We can write equation (\ref{eq:pn13}) in terms of dimensionless variables
$m'_{av,i}/m'_{av,d}$, $t_{1}/\tau_{3}$, and parameter $\mu_{4}/\mu_{3}$,
by writing it in equivalent form 
\begin{equation}
\tau_{3}e^{-t_{2}/\tau_{3}}\left(\frac{m'_{av,i}}{m'_{av,d}}-\left(\frac{m'_{av,i}}{m'_{av,d}}-1\right)e^{t_{1}/\tau_{3}}\right)-\frac{\mu_{4}}{\mu_{3}}\frac{m'_{d}}{m'_{av,d}}t_{2}\cong\tau_{3}+\frac{\mu_{4}}{\mu_{3}}\left(\frac{m'_{i}}{m'_{av,d}}-\frac{m'_{d}}{m'_{av,d}}\right)t_{1}\label{eq:pn15}
\end{equation}
The above model depends also on variables $m'_{i}/m'_{av,d}$ and
$m'_{d}/m'_{av,d}$. Recall that these are ratios of unweighted rates
of change and weighted rates of change, i.e. weighted by $e^{t/\tau_{3}}$.
It is shown later that $m'_{d}/m'_{av,d}$ is approximately constant
across scenarios (also see Figure 3). 

Figure 2 plots the time-delay between peak emissions and concentrations
versus the dimensionless variable $m'_{av,i}/m'_{av,d}$, for the
emissions scenarios plotted in Figure 1. The time-delay increases
with the absolute value of this ratio. Furthermore, for scenarios
with relatively short e-folding mitigation timescale, less than about
40 years, the relationship is approximately linear for the family
of scenarios considered here. 

\begin{figure}
\includegraphics{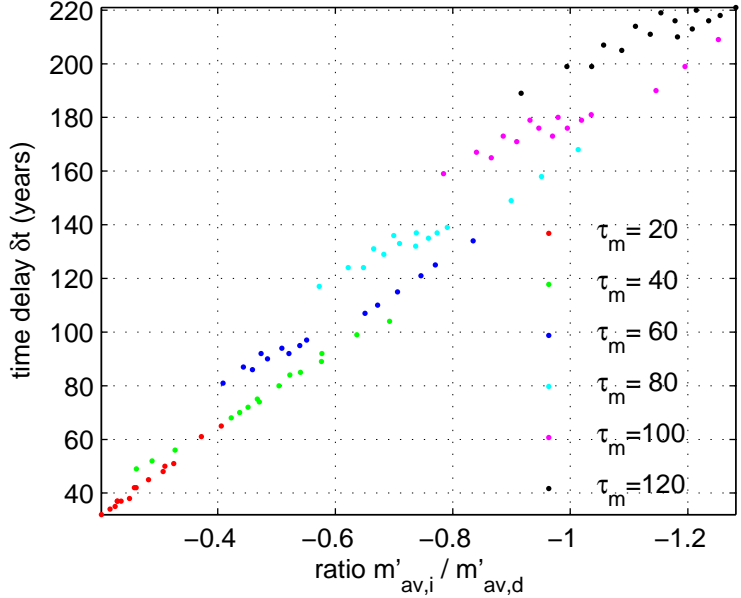}

\protect\caption{Time-delay $\delta t$ between peak emissions and concentrations versus
the ratio of the weighted-averaged rate of increase and decrease of
emissions, defined in equations (\ref{eq:pn6}) and (\ref{eq:pn7})
respectively. The time delay increases with the absolute value of
this ratio. Note the reversed axis in the abscissa. For short e-folding
mitigation timescales, the relationship is approximately linear. Results
are shown for the scenarios in Figure 1, and colors in the plot correspond
to the colors in Figures 1a-d. }
\end{figure}

\subsection{Influence of mitigation parameters, and importance of e-folding mitigation
timescale}

Here we consider the effects of parameters of our specific emissions
model on the rates $m'_{av,i}$, $m'_{av,d}$ , $m'_{i}$ and $m'_{d}$
, and thereby on the solution to equation (\ref{eq:pn15}). This discussion
is of more general interest beyond this particular model because the
parameters - the GGP growth rate, the e-folding mitigation timescale
at which emissions intensity decreases, and the time to stabilization
of GGP - can be easily interpreted. Considering first the rate of
change of emissions during its increasing phase
\begin{equation}
m_{i}'=\frac{\int_{0}^{t_{1}}m'\left(z\right)dz}{\int_{0}^{t_{1}}dz}=\frac{m\left(t_{1}\right)}{t_{1}}\label{eq:p16}
\end{equation}
Approximating $\left(1+r\right)^{t}\cong e^{rt}$ the formula for
emissions becomes $m\left(t_{1}\right)=m_{0}e^{r\left(t_{1}-t_{0}\right)-\left(t_{1}-t_{0}\right)/\tau_{m}}$.
If $r>1/\tau_{m}$ then $t_{1}=t_{g}$, i.e. emissions peaks when
GGP is maximum. Otherwise $t_{1}=t_{0},$ the present. In the first
case $m\left(t_{1}\right)=m_{0}e^{\left(r-\frac{1}{\tau_{m}}\right)\left(t_{g}-t_{0}\right)}$
and $m_{i}'=m_{0}e^{\left(r-\frac{1}{\tau_{m}}\right)\left(t_{g}-t_{0}\right)}/t_{g}$,
whereas in the second case $m\left(t_{1}\right)=m_{0}$ and $m_{i}'=m_{0}/t_{0}$. 

In the following we consider only the case where $r>1/\tau_{m}$ because
global emissions of CO\textsubscript{2} are expected to continue
to increase for a while. For its decreasing phase 
\begin{equation}
m_{d}'=\frac{\int_{t_{1}}^{t_{2}}m'\left(z\right)dz}{\int_{t_{1}}^{t_{2}}dz}=\frac{m\left(t_{2}\right)-m\left(t_{1}\right)}{t_{2}-t_{1}}\label{eq:p17}
\end{equation}
Then $m\left(t_{2}\right)=m\left(t_{1}\right)e^{-\frac{t_{2}-t_{1}}{\tau_{m}}}$
so that $m_{d}'=-m\left(t_{1}\right)\left(1-e^{-\frac{t_{2}-t_{1}}{\tau_{m}}}\right)/\left(t_{2}-t_{1}\right)$,
which becomes $-m_{0}e^{\left(r-\frac{1}{\tau_{m}}\right)\left(t_{g}-t_{0}\right)}\left(1-e^{-\frac{t_{2}-t_{g}}{\tau_{m}}}\right)/\left(t_{2}-t_{g}\right)$
. 

For the weighted rate of increase of emissions between $0$ and $t_{1}$
, we decompose the integral in the numerator of equation (\ref{eq:pn6})
into that between $0$ and $t_{0}$ and between $t_{0}$ and $t_{1}$.
Using $m'\left(t\right)=\left(r-\frac{1}{\tau_{m}}\right)m_{0}e^{\left(r-\frac{1}{\tau_{m}}\right)\left(t-t_{0}\right)}$
between $t_{0}$ and $t_{1}$ we obtain 
\begin{equation}
m'_{av,i}=\frac{m'_{av0}+m_{0}\frac{r-\frac{1}{\tau_{m}}}{r-\frac{1}{\tau_{m}}+\frac{1}{\tau_{3}}}\left(e^{\left(r-\frac{1}{\tau_{m}}\right)\left(t_{g}-t_{0}\right)+\frac{t_{g}}{\tau_{3}}}-e^{\frac{t_{0}}{\tau_{3}}}\right)}{\tau_{3}\left(e^{t_{g}/\tau_{3}}-1\right)}\label{eq:p18}
\end{equation}
 where $m'_{av0}=\int_{0}^{t_{0}}e^{z/\tau_{3}}m'\left(z\right)dz$.
Likewise using $m'\left(t\right)=-\frac{m_{0}}{\tau_{m}}e^{r\left(t_{g}-t_{0}\right)-\frac{t-t_{0}}{\tau_{m}}}$
between $t_{1}$ and $t_{2}$
\begin{equation}
m'_{av,d}=-\frac{m_{0}e^{\left(r-\frac{1}{\tau_{m}}\right)\left(t_{g}-t_{0}\right)}\left(e^{\frac{t_{g}}{\tau_{3}}}-e^{-\frac{t_{2}-t_{g}}{\tau_{m}}+\frac{t_{2}}{\tau_{3}}}\right)}{\left(\tau_{3}-\tau_{m}\right)\left(e^{t_{2}/\tau_{3}}-e^{t_{g}/\tau_{3}}\right)}\label{eq:p19}
\end{equation}

Let us compare the rates $m_{d}'$ and $m'_{av,d}$ by considering
ratio 
\begin{equation}
\frac{m'_{av,d}}{m_{d}'}=\frac{e^{\frac{t_{g}}{\tau_{3}}}\left(1-e^{-\frac{t_{2}-t_{g}}{\tau_{m}}+\frac{t_{2}-t_{g}}{\tau_{3}}}\right)}{1-e^{-\frac{t_{2}-t_{g}}{\tau_{m}}}}\frac{t_{2}-t_{g}}{\left(\tau_{3}-\tau_{m}\right)e^{\frac{t_{g}}{\tau_{3}}}\left(e^{\frac{t_{2}-t_{g}}{\tau_{3}}}-1\right)}\label{eq:p20}
\end{equation}
which, assuming $1/\tau_{m}\gg1/\tau_{3}$ because the mitigation
timescale is generally much shorter than this time-constant of CO\textsubscript{2},
simplifies to
\begin{equation}
\frac{t_{2}-t_{g}}{\left(\tau_{3}-\tau_{m}\right)\left(e^{\frac{t_{2}-t_{g}}{\tau_{3}}}-1\right)}\label{eq:p21}
\end{equation}
In case $\frac{t_{2}-t_{g}}{\tau_{3}}\ll1$ the exponential term in
equation (\ref{eq:p21}) simplifies to $1+\frac{t_{2}-t_{g}}{\tau_{3}}$,
in which case the ratio approximates to
\begin{equation}
\frac{m'_{av,d}}{m_{d}'}\cong\frac{\tau_{3}}{\tau_{3}-\tau_{m}}\label{eq:p22}
\end{equation}
which, for cases where $\tau_{3}\gg\tau_{m}$, approximates to 1.
Hence the value of weighted average $m'_{av,d}$, while not exactly
equal that of the average rate of change of emissions $m_{d}'$ in
its decreasing phase, closely approximates the latter, especially
in cases of rapid mitigation when, in addition, the time-delay between
peak emissions and concentrations is short compared to time-constant
$\tau_{3}$. A near-equality between these two variables, weighted
and unweighted, is seen in Figure 3b. Similarly Figure 3a shows the
relationship between weighted and unweighted rates of change for the
phase when emissions are increasing. While departure from equality
is larger, the relationship is still approximately linear. 

\begin{figure}
\includegraphics{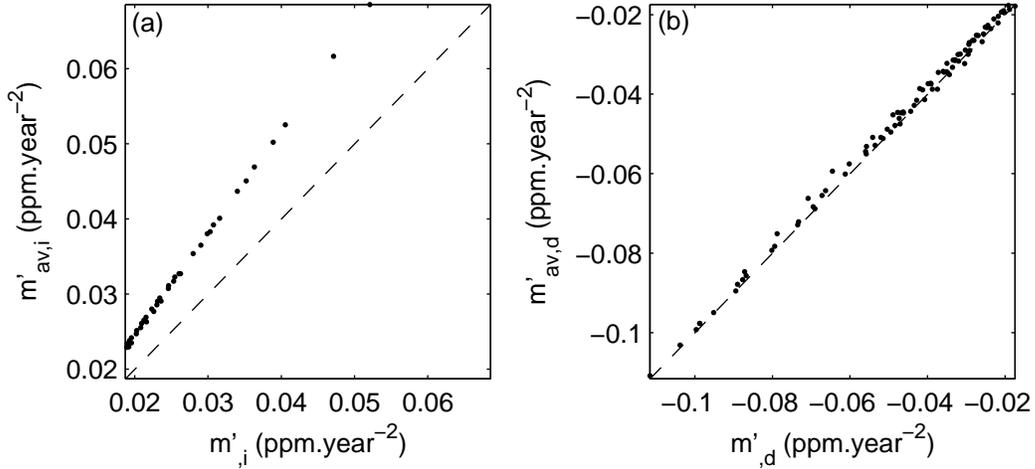}

\protect\caption{Plots of weighted (by $e^{t/\tau_{3}}$) versus unweighted rate of
change of emissions: (a) increasing phase; (b) decreasing phase. Dashed
lines show where abscissa and ordinate are equal. An approximate linear
relationship holds between the two sets of variables.}

\end{figure}

Although these weighted and unweighted rates are not the same, because
of their similarities we can examine ratio $m_{i}'/m_{d}'$ , which
is more tractable, to understand qualitatively what controls the behavior
of ratio $m'_{av,i}/m'_{av,d}$. The former ratio can be written as
\begin{equation}
\frac{m_{i}'}{m_{d}'}=-\frac{t_{2}-t_{g}}{t_{g}}\frac{1}{1-e^{-\frac{t_{2}-t_{g}}{\tau_{m}}}}\label{eq:p23}
\end{equation}
A longer time $t_{2}$ to the concentration peak by itself can increase
the above ratio, because the graph of emissions is convex in its decreasing
phase, so that its slope is decreasing. The influence of time $t_{g}$
to peak emissions is weak because of two countervailing influences:
shorter $t_{g}$ increases the magnitude of the numerator as well
as that of the denominator above. The main influence on this ratio
is that of mitigation timescale $\tau_{m}$ . Short $\tau_{m}$ decreases
the magnitude of this ratio. 

The strong influence of the mitigation timescale on the ratio $m'_{av,i}/m'_{av,d}$
is seen in Figure 4. Short mitigation timescale, corresponding to
rapid mitigation of emissions intensity, is therefore essential to
limit the absolute magnitude of this ratio and therefore assure a
short time-delay. 

Note that the ratio in equation (\ref{eq:p23}) does not depend on
the GGP's growth rate. While this rate influences peak emissions of
CO\textsubscript{2}, it affects the average growth rate and decrease
of emissions in the same manner and hence is not a factor in this
ratio and consequently in the time-delay between peak emissions and
concentrations. 

\begin{figure}
\includegraphics{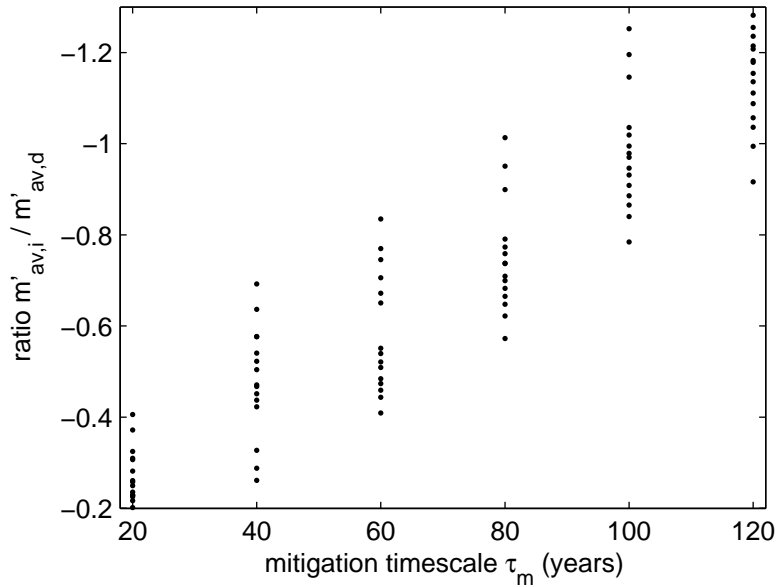}

\protect\caption{Ratio $m'_{av,i}/m'_{av,d}$ versus e-folding mitigation timescale.
Note the reversed axis on the ordinate. Short mitigation timescale
decreases the absolute value of this ratio.}

\end{figure}

\subsection{Carbon cycle uncertainties}

As indicated earlier the carbon-cycle parameters affecting the time
to the concentration peak are the multi-century time-constant $\tau_{3}$
and the ratio $\mu_{4}/\mu_{3}$, describing the ratio of the impulse
response of atmospheric CO\textsubscript{2} from the infinite time-constant
$\tau_{4}$ and the long time-constant $\tau_{3}$ respectively. Figure
5 plots the influence of these parameters on the time-delay. These
are the uncertainties in the carbon cycle that must be constrained
in order to constrain forecasts of the timing of peak concentrations
of CO\textsubscript{2}. 

\begin{figure}
\includegraphics{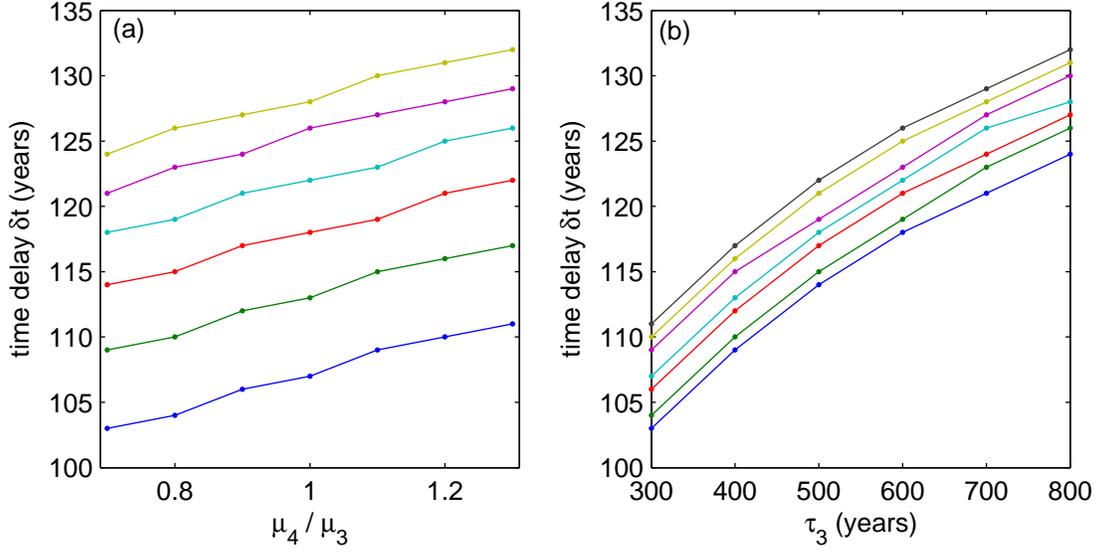}

\protect\caption{The main carbon cycle parameters affecting the time-delay between
peak CO\protect\textsubscript{2} emissions and concentrations: (a)
relation between time-delay and $\mu_{4}/\mu_{3}$ describing the
ratio of the impulse response of atmospheric CO\protect\textsubscript{2}
from the infinite time-constant $\tau_{4}$ and the long time-constant
$\tau_{3}$ respectively; (b) relation between time-delay and long
time-constant $\tau_{3}$. In each panel, different curves correspond
to different fixed values of the other parameter. The emissions scenario
used has $t_{g}-t_{0}=60$ years, $r=0.015$ \%/year, and $\tau_{m}=50$
years. }
\end{figure}

\section{Conclusions and Discussion}

The results presented here are based on approximating the linear carbon
cycle model with four time-constants of \citet{Joos2013}, with one
time-constant being infinite because a fraction of emitted CO\textsubscript{2}
persists for a very long time (\citet{Archer2005,Archer2008}). We
identified the main factors governing the time-delay between peak
CO\textsubscript{2} emissions and concentrations. 

On the emissions front, the main factor is the e-folding timescale
with which the mitigation of emissions intensity of GGP (\textquotedbl{}decarbonization\textquotedbl{})
occurs. This can be viewed as the inverse of the corresponding mitigation
rate (\citet{Seshadri2015a}). Short decarbonization timescale leads
to short time-delay between emissions and concentration peaks. Therefore
achieving decarbonization rapidly is important to achieving an early
peak in CO\textsubscript{2} concentrations. 

The time-delay between peak emissions and concentrations is not sensitive
to the time to peak emissions. However an early emissions peak will
facilitate an early concentration peak. 

The growth rate of economic output is an important factor in peak
emissions. However as discussed here it does not affect the time delay
between peak emissions and concentrations, because it has the same
effect on the rate of increase of emissions and the rate of decrease
of emissions, whose ratio governs the time-delay. Therefore in this
model where peak emissions corresponds to where GGP stabilizes, the
timing of peak concentrations is not affected by this growth rate.
Nevertheless it is obviously important in considering the magnitude
of peak emissions, and faster economic growth has to be accompanied
by more rapid decarbonization. 

The important non-dimensional parameter in our discussion has been
the ratio of the rate of increase of emissions and the rate of decrease
of emissions. Limiting the magnitude of this ratio will help achieve
an early peak, and this can be accomplished by keeping the mitigation
timescale short. 

With respect to the atmospheric cycle of CO\textsubscript{2}, the
influential parameters are the long but finite time-constant $\tau_{3}$
that occurs on century-scales, and the factor $\mu_{4}/\mu_{3}$ describing
the ratio of the impulse response function of atmospheric CO\textsubscript{2}
from the infinite time-constant $\tau_{4}$ and the long time-constant
$\tau_{3}$ respectively. The time-delay increases with these parameters.
The short time constants $\tau_{1}$ and $\tau_{2}$ occurring on
decadal scales or less play a small role in the long-term dynamics
of atmospheric CO\textsubscript{2}, and uncertainties in their values
are correspondingly less important for forecasting the concentration
peak. 

In summary it is important to constrain these carbon cycle parameters,
in addition to achieving an early mitigation peak, as well as implementing
decarbonization on short timescales.

\section*{Acknowledgments}

This research has been supported by Divecha Centre for Climate Change,
Indian Institute of Science. The author thanks G Bala, Michael MacCracken,
and J Srinivasan for helpful discussion.

\bibliographystyle{agufull08}
\bibliography{TimeDelay_refs}

\end{document}